%% file: main.tex
\documentclass[hidelinks, conference]{IEEEtran}
\IEEEoverridecommandlockouts
\PassOptionsToPackage{bookmarks={false}}{hyperref} 

\usepackage[a-1b]{pdfx}
\usepackage{color}
\usepackage{xcolor}
\usepackage{soul}
\usepackage{url}
\usepackage[utf8]{inputenc}
\usepackage{amsthm}
\usepackage{amsmath}
\usepackage{amsthm}
\usepackage{amssymb}
\usepackage{graphicx}
\usepackage{epstopdf}
\usepackage{wrapfig}
\usepackage{algorithmic}
\usepackage{dirtytalk}
\usepackage{algorithm}
\usepackage{mathrsfs}
\usepackage{mathtools}
\usepackage{tabulary}
\usepackage{booktabs}
\usepackage{caption}
\usepackage{xprintlen}
\usepackage{multirow}

\graphicspath{{./}{./files/figures_eps/}}

\PassOptionsToPackage{bookmarks={false}}{hyperref}

\usepackage{subcaption}

\def\BibTeX{{\rm B\kern-.05em{\sc i\kern-.025em b}\kern-.08em
   T\kern-.1667em\lower.7ex\hbox{E}\kern-.125emX}}

\newcommand{\comment}[1]{ }
\newcommand{\proposed}{{\tt{EPS}}} 
\newcommand{\framework}{{\tt{EPS-CNN}}} 
\newcommand{\iqcnn}{{\tt{IQ-CNN}}} 

\newcommand\subparagraph{%
  \@startsection{subparagraph}{0}
  {\parindent}
  {0ex \@plus 0ex \@minus 0ex}
  {-1em}
  {\normalfont\normalsize\bfseries}}
\makeatother

\begin{document}




\title{EPS: Distinguishable IQ Data Representation for Domain-Adaptation Learning of Device Fingerprints}
\title{Domain-Agnostic Signal Representation for Hardware Fingerprinting-Based Zero Trust Authentication}
\title{Domain-Agnostic Hardware Fingerprints for Zero-Trust Wireless Device Identity Identification}
\title{Domain-Agnostic Hardware Fingerprinting-Based Device Identifier for Zero-Trust IoT Architecture}
\title{Domain-Agnostic Hardware Fingerprinting-Based Device Identifier for Zero-Trust IoT Security}
 
\author{\IEEEauthorblockN{Abdurrahman Elmaghbub}
\IEEEauthorblockA{Oregon State University\\
elmaghba@oregonstate.edu}
\and
\IEEEauthorblockN{Bechir Hamdaoui}
\IEEEauthorblockA{Oregon State University\\
hamdaoui@oregonstate.edu}
\thanks{This work is supported in part by NSF/Intel Award No. 2003273.}
}

\maketitle
\thispagestyle{plain}
\pagestyle{plain}

\begin{abstract}
\input{0-abstract}
\end{abstract}


\section{Introduction}
\label{sec:into}
\input{1-introduction}

\section{Understanding the Impact of Oscillator's Frequency Inaccuracy on IQ Signal Behavior}
\label{sec:motiv}
\input{3-motivation}

\section{Distinguishable IoT Device Identification Through Novel RF Signal Representation}
\label{sec:eps}
\input{4-eps.tex}

\section{\proposed-Based Fingerprinting Framework for Domain-Agnostic Device Identification}
\label{sec:proposed}
\input{5-proposed}

\section{Conclusion}
\label{sec:conc}

\input{8a-conclusion}

\bibliographystyle{IEEEtran}
\bibliography{IEEEexample}

\begin{IEEEbiographynophoto}{Abdurrahman Elmaghbub}~received the B.S. degree with summa cum laude, and MS in Electrical and Computer Engineering from Oregon State University in 2019, and 2021, respectively, and is currently pursuing his Ph.D. degree in the School of Electrical Engineering and Computer Science at Oregon State University. His research interests are in the area of wireless communication and networking with a current focus on applying deep learning to wireless device classification.
\end{IEEEbiographynophoto}

\begin{IEEEbiographynophoto}
    {Bechir Hamdaoui}~is a Professor in the School of Electrical Engineering and Computer Science at Oregon State University. He received M.S. degrees in both ECE (2002) and CS (2004), and the Ph.D. degree in ECE (2005) all from the University of Wisconsin-Madison. 
His general interests are in theoretical and experimental research that enhances the cybersecurity and resiliency of future intelligent networked systems. He is the Founding Director of the NetSTAR Laboratory at Oregon State University.
Dr. Hamdaoui and his team have won several awards, including the ISSIP 2020 Distinguished Recognition Award, the 2009 NSF CAREER Award, the ICC 2017 Best Paper Award, and the 2016 EECS Outstanding Research Award. He serves/served as an Associate Editor for several IEEE journals and magazines and chaired \& organized many IEEE/ACM conference symposia \& workshop programs. He served as a Distinguished Lecturer for the IEEE Communication Society in 2016 and 2017 and served as the Chair \& Co-chair of the IEEE Communications Society's Wireless Technical Committee (WTC) from January 2019 until December 2022.
\end{IEEEbiographynophoto}
\end{document}

%% file: 0-abstract.tex
{Next-generation networks aim for comprehensive connectivity, interconnecting humans, machines, devices, and systems seamlessly. This interconnectivity raises concerns about privacy and security, given the potential network-wide impact of a single compromise. To address this challenge, the Zero Trust (ZT) paradigm emerges as a key method for safeguarding network integrity and data confidentiality. This work introduces EPS-CNN, a novel deep-learning-based wireless device identification framework designed to serve as the device authentication layer within the ZT architecture, with a focus on resource-constrained IoT devices. At the core of EPS-CNN, a Convolutional Neural Network (CNN) is utilized to generate the device identity from a unique RF signal representation, known as the Double-Sided Envelope Power Spectrum (EPS), which effectively captures the device-specific hardware characteristics while ignoring device-unrelated information. 
Experimental evaluations show that the proposed framework achieves over 99\%, 93\%, and 95\% of testing accuracy when tested in same-domain (day, location, and channel), cross-day, and cross-location scenarios, respectively. 
%
Our findings demonstrate the superiority of the proposed framework in enhancing the accuracy, robustness, and adaptability of deep learning-based methods, thus offering a pioneering solution for enabling ZT IoT device identification.}

%% file: 1-introduction.tex
In the ever-evolving landscape of intelligent and interconnected wireless systems, the comprehensive connectivity, enabled by massive Internet of Things (IoT) networks, and the surge in sophisticated cyber-attacks have highlighted the pressing need for a revolutionary approach to network security. Traditional approaches, built on the concept of a secure physical perimeter, are falling short of the increasingly demanding security measures as the key assumptions of these models no longer hold \cite{ward2014beyondcorp}. In response to this growing concern, the Zero Trust (ZT) model has emerged as a transformative paradigm, redefining the fundamental principles of network security \cite{stafford2020zero}. Rooted in the philosophy of ``never trust, always verify," the ZT model advocates for a proactive and dynamic approach to security, where trust is no longer assumed, but continuously verified throughout the entire network environment, and resource access is granted solely based on device and user credentials, irrespective of the user's network location. The new paradigm offers special attention to the identity authentication process as any misstep at this stage can jeopardize the integrity of the entire system. 
 The foundational cornerstone within the ZT paradigm centers on the robust necessity for unequivocal device identity. This imperative encompasses an array of requisites including the hardware root of trust, passwordless authentication, renewable credentials, and device registry \cite{hunt2017seven}. In accordance with these tenets, highly-secured devices possess a cryptographically-backed, unique, and unforgeable ``onboarding'' identity that is inseparable from the hardware and managed by an embedded security processor. They also enjoy a passwordless authentication process leveraging digital certificates signed and verified using private and public cryptographic keys. These certificates extend their efficacy to provide renewable and operational credentials for continuously secured operations. Lastly, it is crucial to have a device registry that stores the core attributes of the devices to facilitate and audit the access process. These requisites necessitate the incorporation of pivotal components including cryptographic engines, security processors, and secure storage \cite{hunt2017seven}. However, this integration poses challenges for the vast number of microcontroller-based IoT devices, constrained by factors such as size, power, or cost limitations. With countless interconnected devices forming integral parts of expanding IoT ecosystems, the compromise of even seemingly innocent devices can catalyze escalating threats, encompassing data pollution, lateral movement, and denial-of-service attacks. Hence, safeguarding each individual device within the network attains paramount significance. 
 
 To bridge this gap, we propose a novel hardware device fingerprinting-based framework, named \framework, that builds a unique and unforgeable hardware-based identity for IoT device authentication. 
Hardware device fingerprinting technology serves as a powerful physical-layer security mechanism, enabling device identification through the extraction of unique device fingerprints embedded in the devices' transmitted signals \cite{sankhe2019oracle, jagannath2022comprehensive}. These fingerprints emerge due to inherent hardware manufacturing imperfections of various {Radio Frequency} (RF) circuitry components 
yielding signal distortions~\cite{elmaghbub2021} that collectively shape distinctive device signatures. 

%
{Deep Learning (DL) stands out as a powerful computational paradigm for embedding intelligence in IoT networks. 
It has been, recently, utilized for device discovery, vulnerability analysis, anomaly detection, and trust-based policy recommendations} \cite{hussain2020machine}. 
Although DL-based hardware device fingerprinting methods have demonstrated promising results in terms of device identification accuracy, 
several studies (e.g., \cite{elmaghbub2021, hamdaoui2022deep, al2020exposing, hanna_wisig_2022}) have revealed that many of these methods do not perform well when the testing data is collected under a {\em domain} that is different from that used during training. Here, the term {\em domain} refers to the network setting/environment (e.g., channel {condition}, device location, etc.) under which data is collected. {This observation can be attributed to the prevalent use of raw In-phase/Quadrature (IQ) data representation as the input to the DL-based device identifiers} \cite{needle}. {However, within the context of fingerprinting, this representation contains an abundance of device-irrelevant information. Consequently, extracting meaningful fingerprints from this raw IQ data becomes akin to finding a needle in a haystack filled with numerous misleading needle-like objects.

To ensure a flawless authentication operation across various domains, the domain-resilient property becomes an indispensable attribute for any device identification framework aspiring to integrate seamlessly into the ZT architecture. Accordingly, our proposed identification framework introduces a novel RF signal representation, double-sided envelope's power spectrum, referred to as \proposed~for short, which significantly enhances the accuracy and robustness of DL-based hardware device fingerprinting methods against domain changes. Our \proposed~representation vividly captures the device's hardware impairments while suppressing device-irrelevant information. Specifically, it closely mirrors the impaired behavior of a key RF hardware component, the oscillator, whose impairment substantially contributes to the device's unique fingerprint and is proven resilient in the presence of variations in time, channel conditions, and/or location. The adoption of the \proposed~representation is pivotal in mitigating the impact of environmental changes on device identification, ensuring the reliability and stability of our approach. 

To generate this \proposed~representation, we extract the outer shape or envelope of the IQ signal, thereby eliminating resultant amplitude offsets, and produce the double-sided envelope's power spectrum of the received burst, which serves as an effective input to machine learning classifiers. Leveraging \proposed, we then propose the \framework~device identification framework, which channels the output of the \proposed~extractor engine into a standard Convolution Neural Network (CNN) to create device-unique identities from received RF signals.}

Through extensive evaluation on a testbed of 15  WiFi devices, we demonstrate the effectiveness of our \framework~device identification framework in real-world scenarios. 
%
Notably, we show that our framework achieves an accuracy of over 99\% in same-domain scenarios (i.e. training and testing are both done on same day/location), and more importantly, consistently sustains an accuracy that exceeds 95\% for cross-location scenarios and 93\% for cross-day scenarios. 
%


The key contributions of this work can be summarized as: 
\begin{itemize}

    \item We propose \proposed, a novel IQ representation that significantly enhances the performance of deep learning-based device identifiers. We demonstrate, through experimentation, the distinguishability and reliability of \proposed~under varying time, channel, and/or location domains. 

    \item We propose \framework, an \proposed-based device fingerprinting framework that substantially enhances the accuracy and robustness of device identification in the presence of varying domains.

    
    \item We assess the effectiveness of \framework~in identifying WiFi devices and showcase an exceptional cross-domain performance in real-world scenarios, achieving an average testing accuracy of 93\% and 95\% respectively for the cross-day and cross-location scenarios.

\end{itemize}

The rest of the paper is organized as follows.  Sec.~\ref{sec:motiv} studies the impact of the oscillator's carrier frequency inaccuracy and offset on the behavior of IQ signals. Sec.~\ref{sec:eps} presents the proposed IQ data representation approach, \proposed. Sec.~\ref{sec:proposed} presents the proposed \framework~identification  framework along with its extensive evaluation under different scenarios. Finally, Sec.~\ref{sec:conc} concludes the paper.

%% file: 3-motivation.tex
As highlighted earlier, recent studies \cite{elmaghbub2021, al2020exposing} have underscored a critical drawback in the performance of DL models that solely rely on IQ samples, revealing their struggles with maintaining consistency across diverse domains. 
Consequently, there arises a pressing need for novel RF signal representations that adeptly capture device-specific impairments. These representations are essential to refine the feature selection process within DL-based fingerprinting techniques and enhance their resilience to shifts in different domains. In this work, we propose a device identification framework based on a robust RF signal representation that accurately captures the impairments of oscillators, {which} serves as the foundation for identity generation. 

To be able to design such efficient representations, it is important to begin by studying the impact of the carrier frequency inaccuracy caused by local oscillators on the behavior of the received IQ signals. 

\subsection{The Carrier Frequency Offset (CFO)}
Local oscillators are essential hardware components in the RF transceiver chain, primarily generating oscillating signals that serve as the foundation for establishing RF communication. Therefore, the accuracy of the oscillator, denoting the frequency offset from the specified target frequency, and its stability, which refers to the frequency dispersion around its operational value over time, are critical in RF applications, as they significantly impact the overall system performance \cite{zhou2008frequency}. Carrier Frequency Offset (CFO) is a hardware impairment that arises due to the mismatch between the receiver's local oscillating frequency and that of the sender. This mismatch often occurs due to various factors, including Doppler shifts, oscillator inaccuracies, or synchronization errors in communication systems \cite{vo2016fingerprinting}. 

\subsection{The Impact of Oscillator's Frequency Inaccuracy}

\begin{figure}
\centering
    \begin{minipage}[t]{\linewidth}  
    \centering 
    \subfloat[The I component of the received IQ signal of four devices]{%
    \includegraphics[width=\linewidth,height=.37\linewidth]{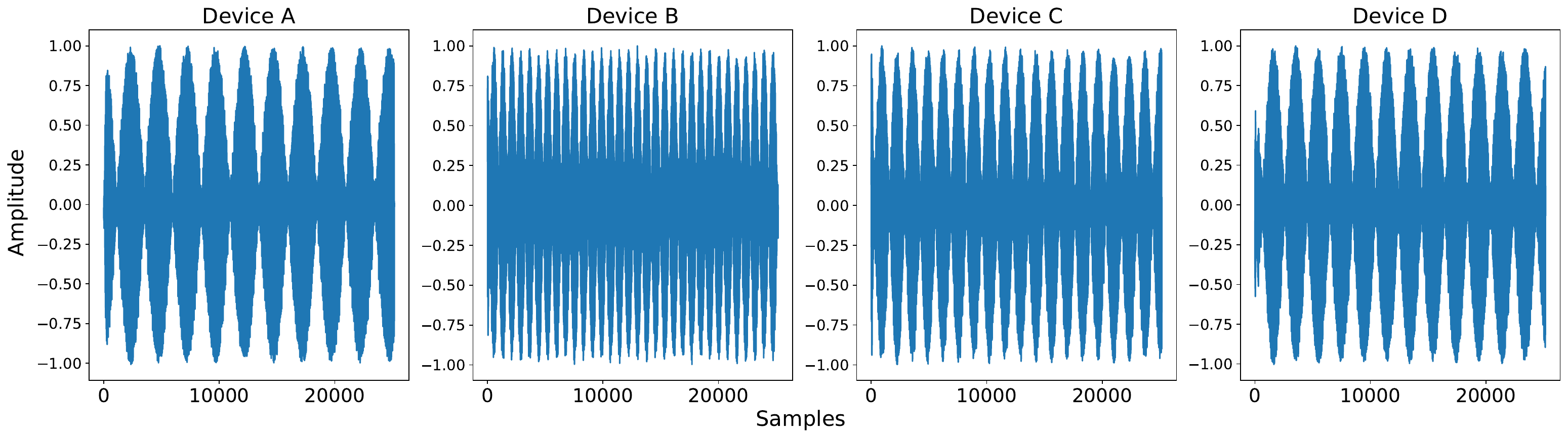} 
     \label{all-devices-I}} 
     \vspace{0.01in}
    \subfloat[The Q component of the IQ signal of four devices]{%
    \includegraphics[width=\linewidth,height=.37\linewidth]{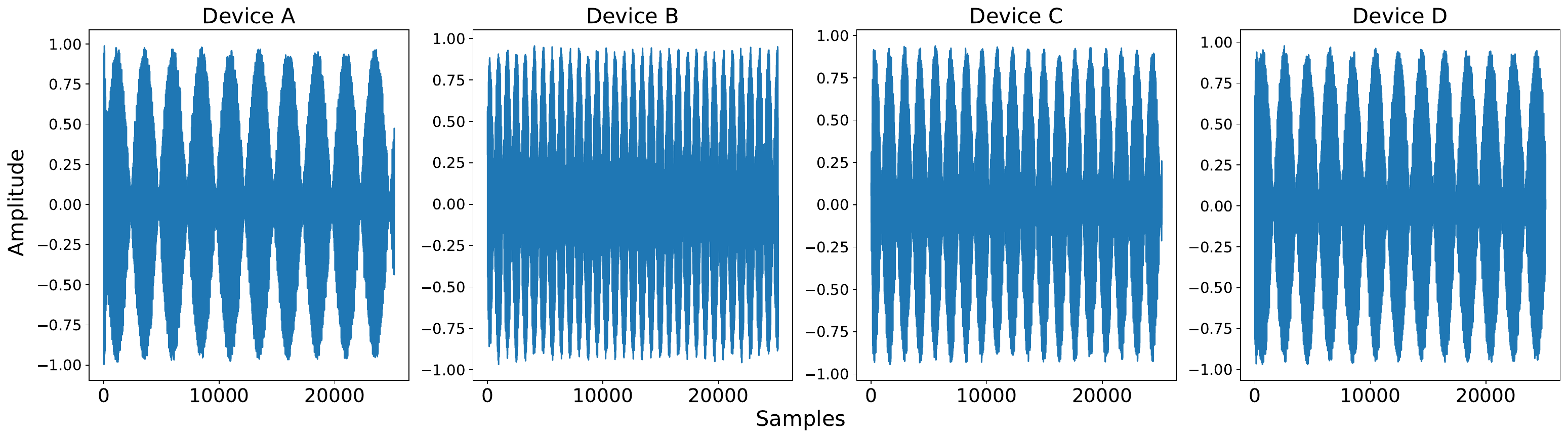}
    \label{all-devices-Q}}
    \caption{The time-domain IQ signal behavior across four different Pycom/IoT devices.} 
    \label{all-devices-abcd} 
    \end{minipage}
\end{figure}

To investigate the impact of the oscillating frequency inaccuracy on the IQ signal behavior, we leveraged our testbed of 15 Pycom/IoT devices (shown in Fig.~\ref{tx}) to analyze the IQ signals collected from multiple different (but identical in hardware) off-the-shelf devices and captured using a USRP B210 receiver.
This is done by having each of the 15 Pycom devices transmit the same IEEE 802.11b WiFi packets after being powered on for 12 minutes to ensure hardware stabilization. We want to emphasize here the importance of waiting until the end of the warm-up period of the devices' hardware before starting the data collection process to ensure robust and consistent measurements \cite{elmaghbub2023eps}. 

We show in Fig.~\ref{all-devices-abcd} the behavior of both the I (in-phase, Fig~\ref{all-devices-I}) and Q (quadrature, Fig.~\ref{all-devices-Q}) components of the time-domain IQ signals captured from four selected devices: Devices A, B, C, and D. We make a couple of key observations from the figure. First, both the I and Q signals showed a ``sinusoidal" pattern in their envelopes, where the envelope of an oscillating signal is the smooth boundary function that outlines the extremes of the signal. 
More importantly, note that the number of ``humps'' of the envelope varied among the devices. 
It is worth mentioning that the reported sinusoidal behaviors of the IQ signal's envelopes are observed across all of the 15 tested Pycom devices, but with each device exhibiting a slightly different number of ``humps''. 

Two pivotal questions now emerge: (i) what underlies the sinusoidal pattern seen in the IQ signal envelope, and (ii) what accounts for the varying count of ``humps" among different devices? 
We proceed to show that the main cause behind such behavior is the CFO (carrier frequency offset) between the Pycom device's oscillating frequency and that of the receiver that exists due to the inaccuracy of the device's local oscillator. 
Specifically, we will next show that the number of humps in the sinusoidal envelope depends on the CFO value. This explains that the reason why different devices exhibit different numbers of humps is that each device presents a different CFO {value}, which varies across devices due to the device's oscillator hardware imperfections incurred during manufacturing.

\begin{figure}
\centering
    \begin{minipage}[t]{\linewidth} 
    \centering 
    \subfloat[CFO = 0 (ideal scenario)]{
    \includegraphics[width=.45\linewidth,height=0.35\linewidth]{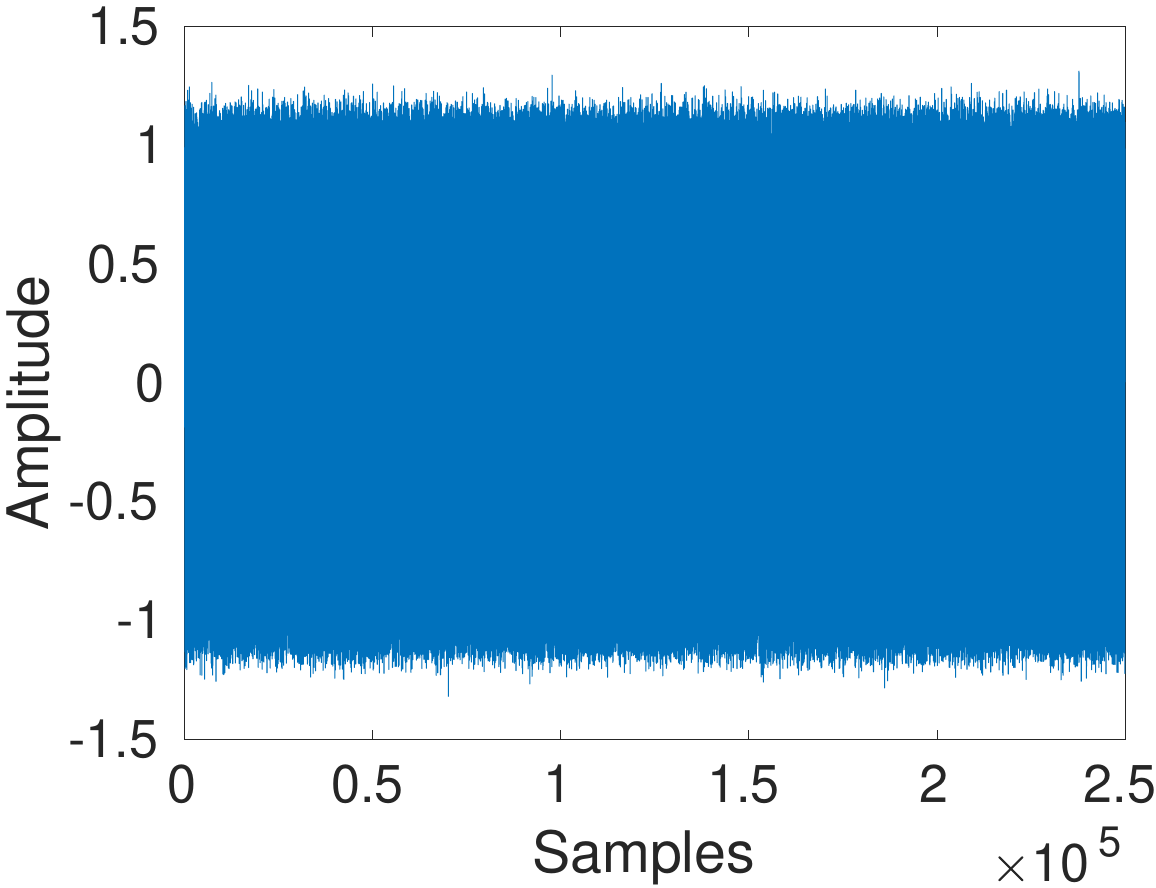} 
     \label{cfo0}}
     \hfill
    \subfloat[CFO = 50 Hz]{
    \includegraphics[width=.45\linewidth,height=0.35\linewidth]{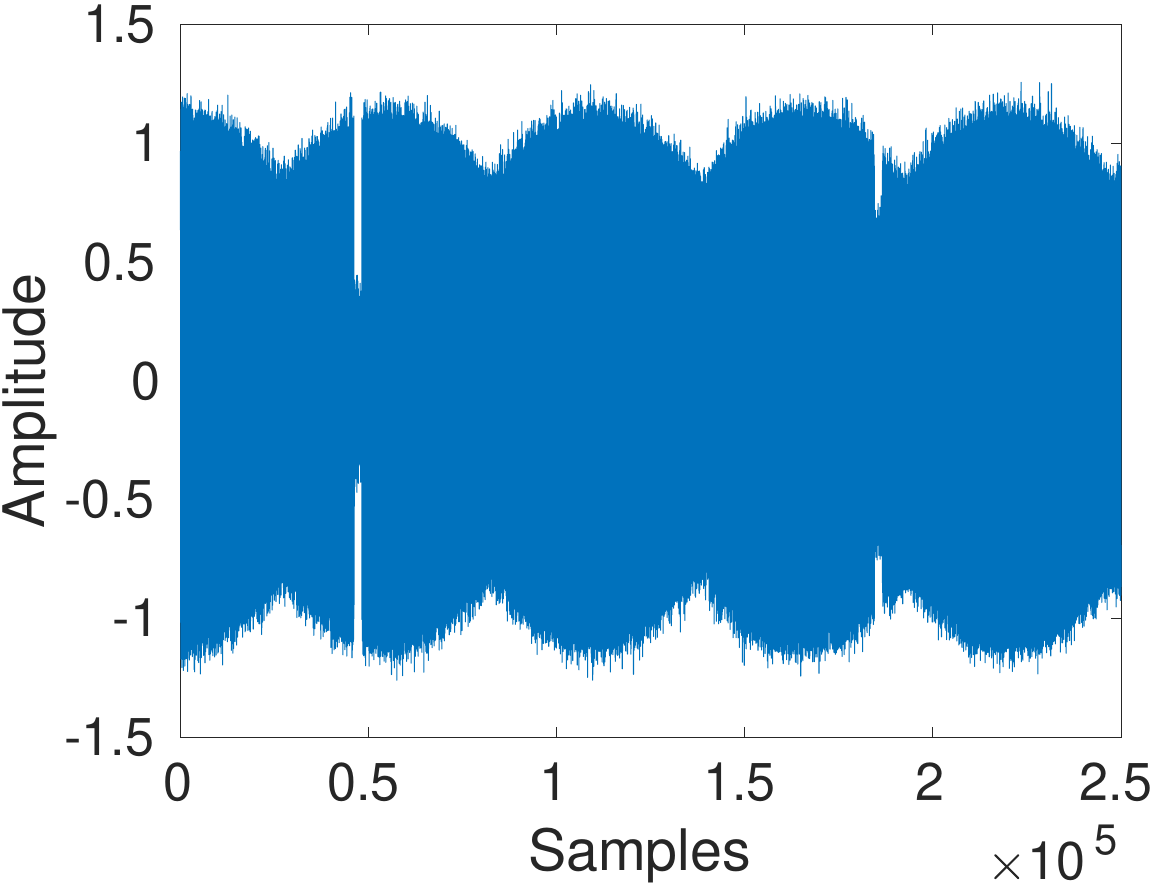}
    \label{cfo50}}\\
    \subfloat[CFO = 100 Hz]{
    \includegraphics[width=.45\linewidth,height=0.35\linewidth]{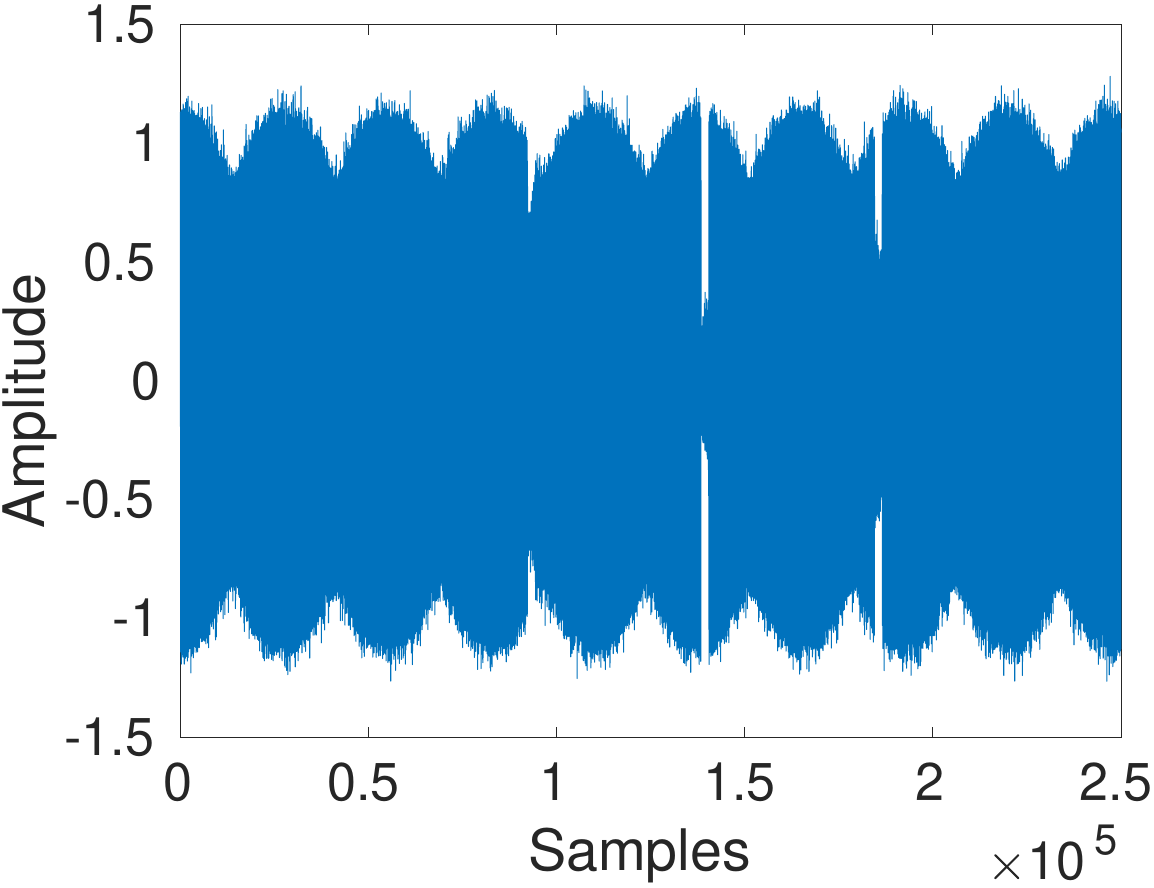}
    \label{cfo100}}
     \hfill
\subfloat[CFO = 200 Hz]{
    \includegraphics[width=.45\linewidth,height=0.35\linewidth]{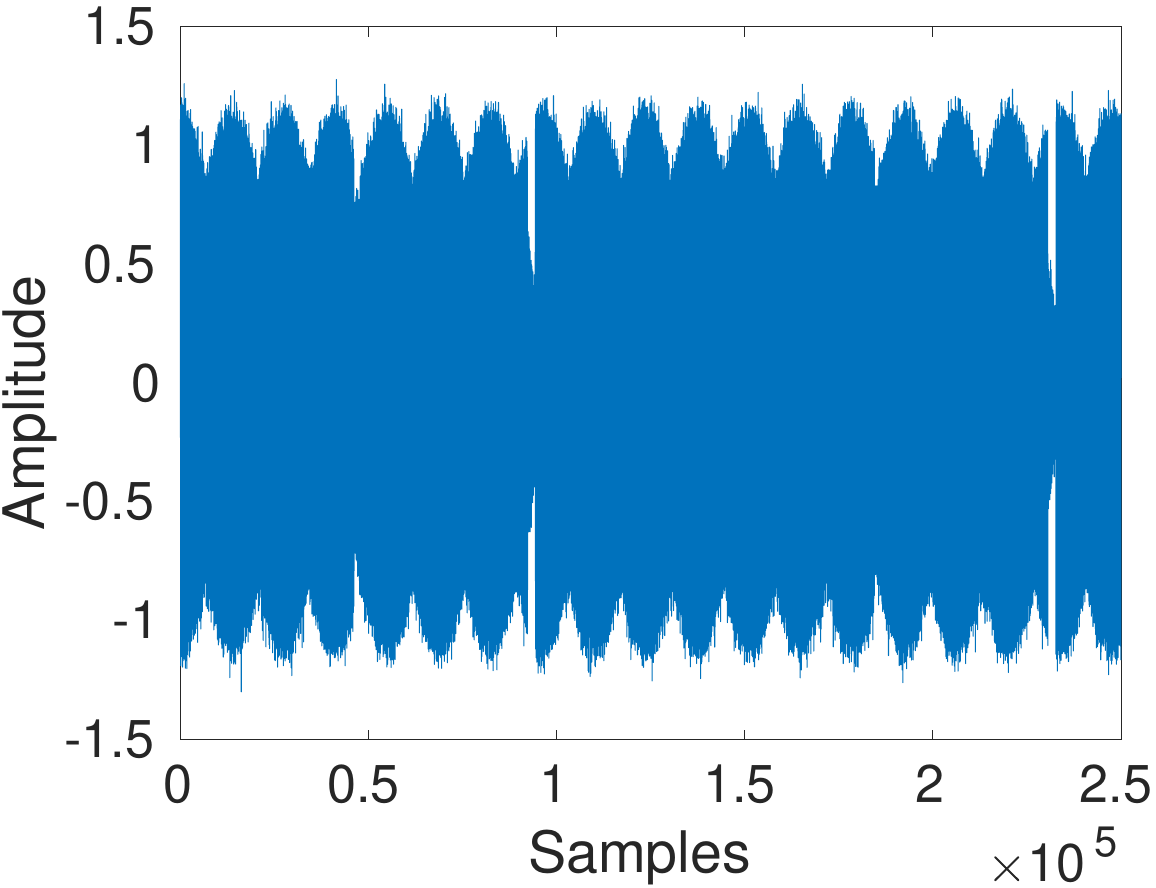}
    \label{cfo200}}
    \caption{Simulated time-domain I signal component.} 
    \label{cfo-matlab} 
    \end{minipage}
\end{figure}


\subsection{What Causes the Observed IQ Envelope Behavior?} 
We utilized MATLAB simulations to validate that the CFO is responsible for the sinusoidal envelope behavior observed in the IQ signal of IoT devices. 
We constructed a complete WLAN 802.11b system which we manipulated to vary the CFO values between the transmitter and receiver. Various CFO impairments were introduced, including scenarios with 0 Hz (ideal device), 50 Hz, 100 Hz, and 200 Hz.
The CFO-impaired transmitted signal is then first passed through an AWGN channel, and then down-converted and sampled by the receiver to generate IQ data samples.
Then, we extracted the real (I) components of the signals and plotted them separately for CFO = 0 in Fig.~\ref{cfo0}, CFO = 50Hz in Fig.~\ref{cfo50}, CFO = 100Hz in Fig.~\ref{cfo100}, and CFO = 200Hz in Fig.~\ref{cfo200}.
The simulated results clearly show the dependency between the CFO values and the number of observed ``humps'' in the I signal's Envelope, and that the CFO is what causes the observed envelope shape. The same trends were observed for the Q signal components as well, but we did not include them here to limit redundancy.

We want to mention that we also experimented with varying other hardware impairments, including IQ imbalance, Phase Noise, and DC offset, but {did not observe} any ``sinusoidal'' behavior of the envelopes. This confirms that other transceiver hardware impairments, though do manifest themselves in other types of distortions, do not yield the Envelope behavior we observed with the CFO impairment.

%% file: 4-eps.tex
In this section, we begin by presenting a novel RF signal representation extracted from the oscillator's envelope shape 
that substantially improves the robustness of device fingerprinting to domain changes and variations. 
Next, we assess the performance of the suggested feature design concerning its capability to effectively serve two main purposes: (i) {\em distinguish between devices} and (ii) {\em withstand} domain changes by maintaining high accuracy performance under varying domains.

\subsection{The Proposed Double-Sided Envelope's Power Spectrum (\proposed) Representation}
\label{sec:eps_gen}
To generate the proposed signal representation, we utilized a sequence of operations that compose the \proposed~generator. The procedure involves creating an analytic signal by initially processing the IQ values of the received frame, denoted as $r(t)$, through an FIR Hilbert transform filter based on the Parks-McClellan algorithm \cite{Park_McClellan}. This filtered output is then scaled by $\sqrt{-1}$ 
and combined with the time-delayed original signal. Incorporating a delay is crucial due to the inherent delay introduced by the FIR filter implementation of the Hilbert transform, equating to half the filter's length. Subsequently, the signal's envelope, denoted as $e(t)$, is derived by computing the absolute value of the analytical signal. This envelope is characterized by a lower frequency compared to the original signal. Consequently, we downsample the signal by a factor of $15$ and then subject it to a lowpass filter to effectively mitigate ringing and smoothen the envelope. Once the envelope is extracted, we center its amplitude around zero before proceeding to generate the corresponding normalized double-sided envelope's power spectrum, i.e., the \proposed~representation, utilizing a power spectrum estimator. The \proposed~representations of various Pycom/IoT devices are visualized in Figure~\ref{fig:unique_spec}. This representation possesses critical attributes in terms of both distinctiveness and robustness, rendering it the suitable foundation for generating strong device identities.
\subsection{\proposed~Distinguishability Across Different Devices}

 \begin{figure}[t] 
    \centering 
    \includegraphics[width=\linewidth]{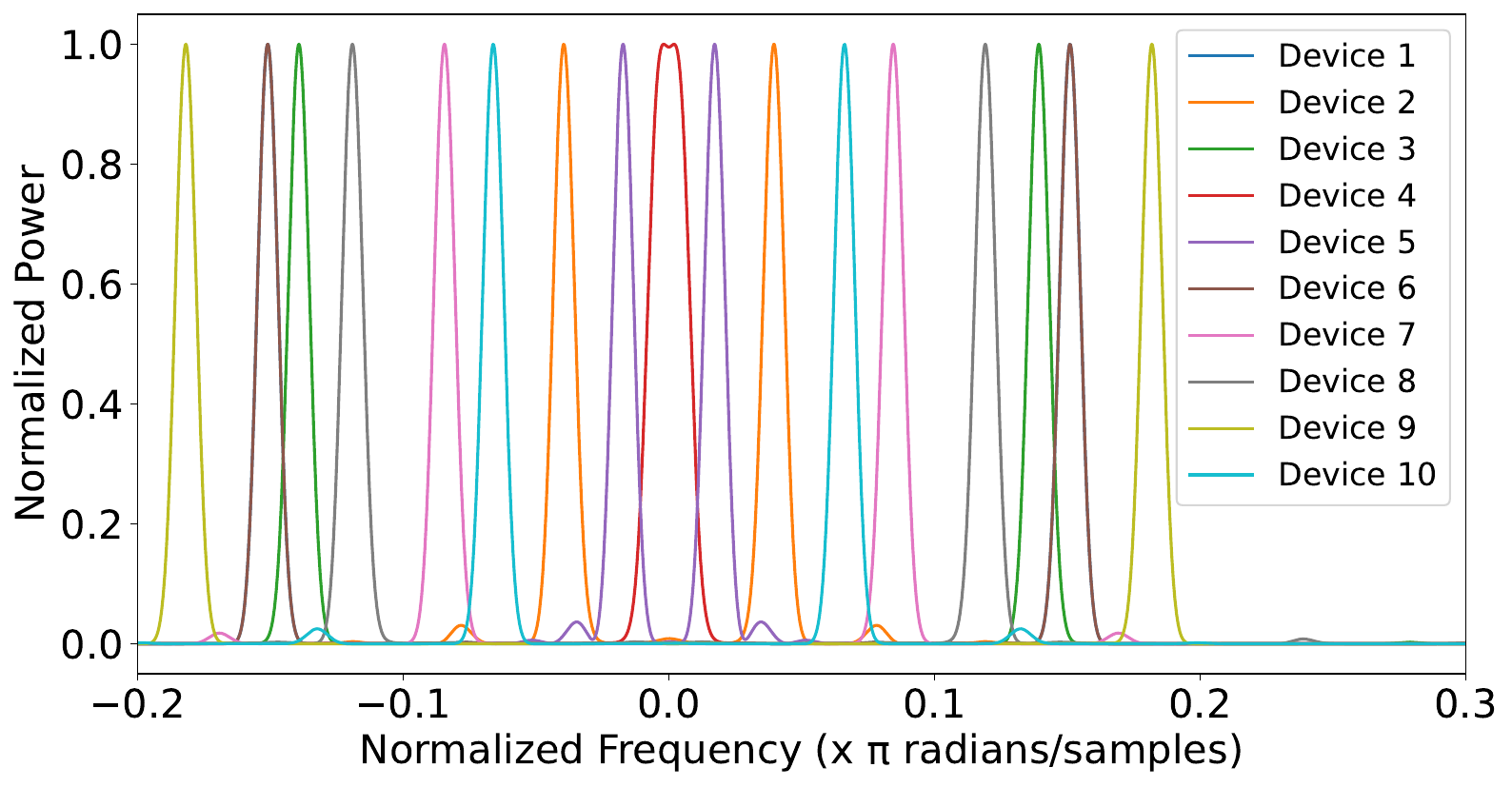} 
    \caption{The \proposed~representation of 10 devices.} 
    \label{fig:unique_spec} 
\end{figure}

In the context of device identification, a signal representation that exhibits distinctive device-specific characteristics is indispensable. The proposed \proposed~representation possesses this property, as it captures the local oscillator's behavior, which is affected by the oscillator's unique hardware impairments. To validate this hypothesis, we conducted an experimental evaluation using a testbed consisting of $15$ Pycom devices, running the IEEE802.11b protocol and a USRP B210 receiver.
Our results depicted in Fig. \ref{fig:unique_spec} reveal that the \proposed~representation is indeed unique for each device, as evidenced by the discernible differences, across the 10 studied devices, in the shape and location of the main sideband and its harmonics. 
This will be further validated through experimental results that are presented later in Sec.~\ref{sec:eval}.

 \subsection{\proposed~Robustness to Domain Changes}

    

\begin{figure}
\subfloat[Channel-Time Domain]{
 \includegraphics[width=\columnwidth]{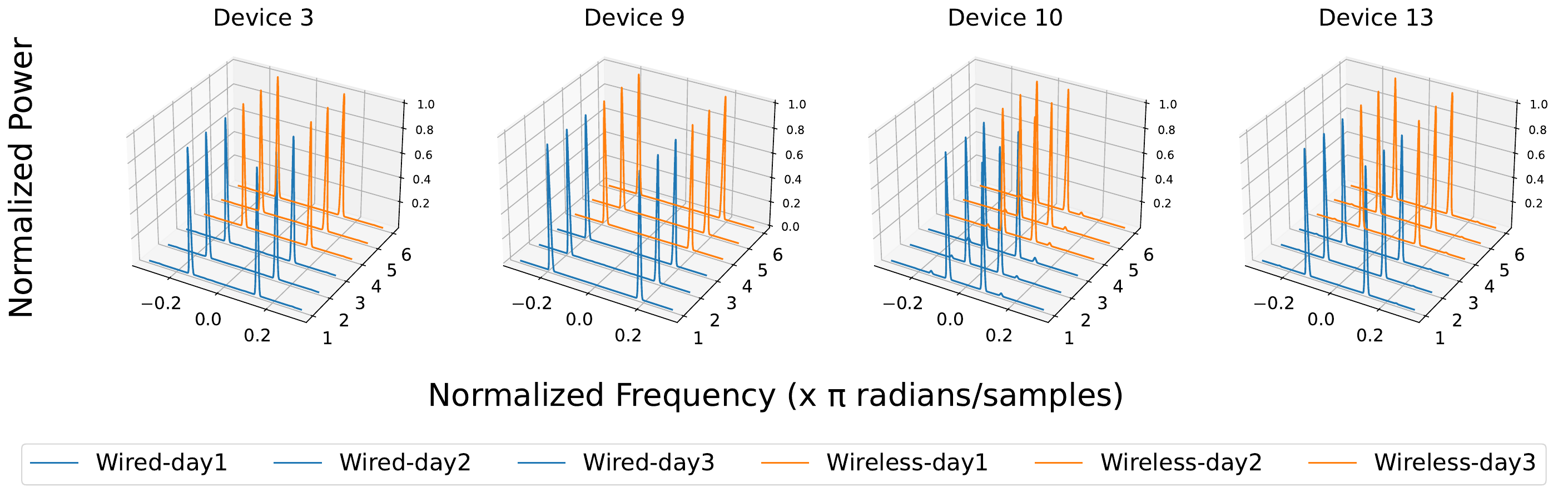}
 \label{channel}}

\subfloat[Location Domain]{
 \includegraphics[width=\columnwidth]{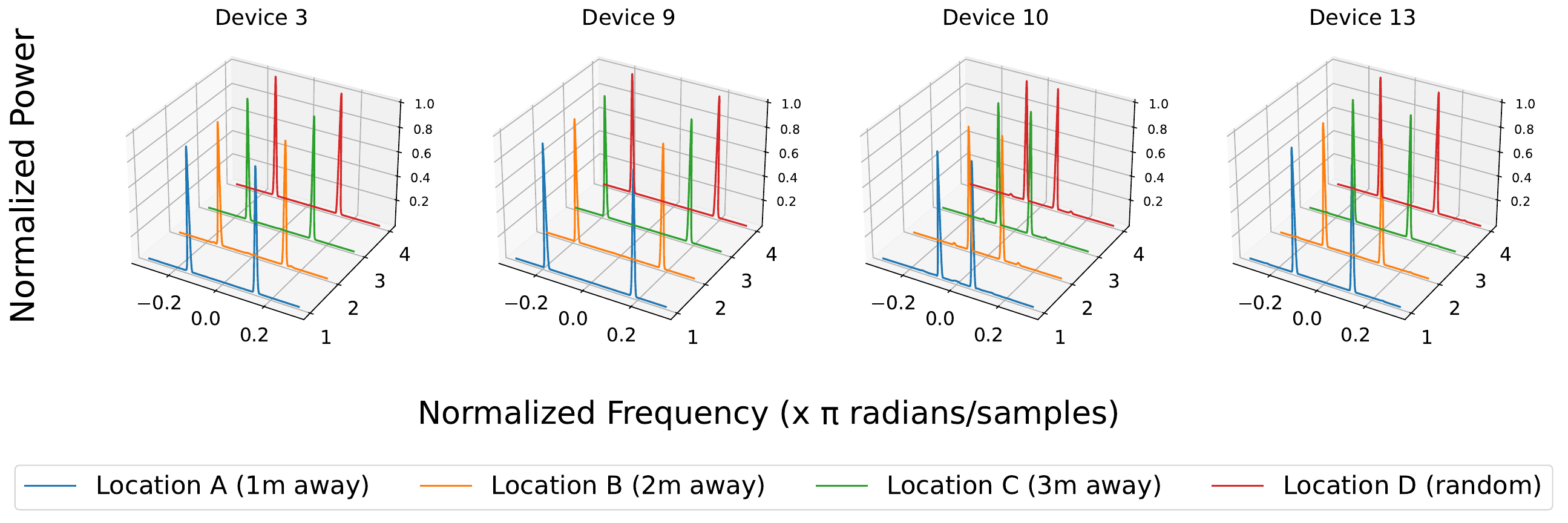}
 \label{location}}

\caption{\proposed~representation resiliency across the channel and location domains}
\label{eps-channel-location}
\end{figure}

Displaying distinctive features that are unique to each device is a crucial aspect, but it alone isn't enough for a representation to serve as a solid foundation for generating device identities. When a representation of a device exhibits random variations as the network's context shifts, it becomes incapable of providing a dependable fingerprint. Consequently, it cannot be employed as an input for a reliable device identification system. Therefore, having demonstrated the distinctiveness of the proposed representation through our testbed, our focus now shifts to evaluating its resilience across three different domains: time, channel, and location.

\subsubsection{Channel-Agnostic Fingerprinting}
To thoroughly investigate how the wireless channel impacts the stability and consistency of \proposed, an extensive experiment was conducted within an indoor environment. In this experimental arrangement, devices were consistently positioned at a fixed distance of 1 meter from the receiver, in both wired and wireless setups. Over three consecutive days, packets were captured and analyzed. The objective of this investigation was to compare the \proposed~representations of packets corresponding to each individual device across both the wired and wireless channels over time, thereby discerning the influence of channel variations over time. Fig.~\ref{channel} presents the graphical representations of the \proposed~representations obtained from four distinct devices under both wired and wireless channel conditions. Notably, the figures demonstrate that the \proposed~representation of each device remains unaltered regardless of the underlying channel characteristics. 
 {This observation is consistent across all 15 devices, providing strong empirical support for the robustness and efficacy of our proposed representation in mitigating sensitivity to channel variations during device identity generation.}


\subsubsection{Location-Agnostic Fingerprinting}
Changing the location and distance between the transmitting devices and the receiver after training can also lead to a drastic drop in performance. To evaluate the robustness of the \proposed~representation to such distance and location changes, we captured data at three different locations with the devices being placed $1$m-away (Location A), $2$m-away (Location B), $3$m-away (Location C), and randomly deployed within a radius of $3$m-away (Location D) from the USRP receiver; this is shown in Fig.~\ref{loc_setup}. 
%
The plots in Fig.~\ref{location} manifest the stability of the \proposed~feature representation over the four studied location scenarios as the signal representations at the four locations completely overlap. 
Our findings confirm the stability of the \proposed~representation in scenarios in which the location, distance, and time of the training and testing sets are different, making the proposed \proposed~representation a more reliable and robust input for wireless device identification.


%% file: 5-proposed.tex

Addressing the device identification challenges arising from domain shifts remains a formidable task. This has hindered the practical adoption of deep learning-aided fingerprinting-based device identification approaches in real-world scenarios. 
In this section, we present our proposed device identification framework, built upon the novel \proposed~feature representation, and show its effectiveness in overcoming such domain-shift challenges by enhancing the resiliency of device identification when faced with changes in the channel condition, device location and/or time of data collection.

\subsection{An Overview of the Proposed EPS-CNN Framework}
Designing an architecture for integrating our \framework~into a zero-trust IoT network involves several components and considerations. At its highest level, the operation flow can be described as follows. New IoT devices are enrolled in the network through a secure enrollment/registration process, during which each IoT device undergoes hardware device fingerprinting. The fingerprinting of these devices, via the \proposed~generator, is initiated by taking the complex-valued IQ representation of a received frame, $r(t)$, as an input (with a dimension of 1x25170) and then processing the I and Q components separately. For each frame, the \proposed~generator first extracts the envelope of the signal, $e(t)$, and then generates the \proposed~representation of the two components: \proposed(I) and \proposed(Q). Refer to Sec.~\ref{sec:eps_gen} for details about the \proposed~representation generation. These two \proposed~representations are then combined into a tensor of size 2x4096, which plays a central role in training the CNN model serving as the device identifier. {Our device identifier encompasses a structured arrangement of six convolutional blocks, which extract the fingerprint from the \proposed~representation, three fully-connected layers with LeakyReLUs in between, and a concluding Softmax layer, which makes determinations regarding the device's identity. Each convolutional block includes 2D-Convolutional, BatchNormalization, LeakyReLU, and MaxPooling layers.}
The device identifier's primary function is to discern patterns within the \proposed~input and make precise predictions about the corresponding device's identity. This completes the enrollment/registration phase. 

Subsequently, when an IoT device attempts to access the network, it goes first through a rogue device detection to confirm its legitimate affiliation with the network. This is followed by identity verification, which employs the hardware fingerprint \proposed~and leverages the device identifier to accurately establish the device's identity. 
For robust security, continuous authentication is implemented on the packet level using the device's received RF signal during its interaction with the network.
Any deviation from the established \proposed~fingerprint triggers an alert for further investigation.

\subsection{Performance Metrics}
\label{sec:eval}



\begin{figure}
     \subfloat[15 Pycom transmitting devices\label{tx}]{%
       \includegraphics[width=0.23\textwidth, height = 0.2\textwidth]{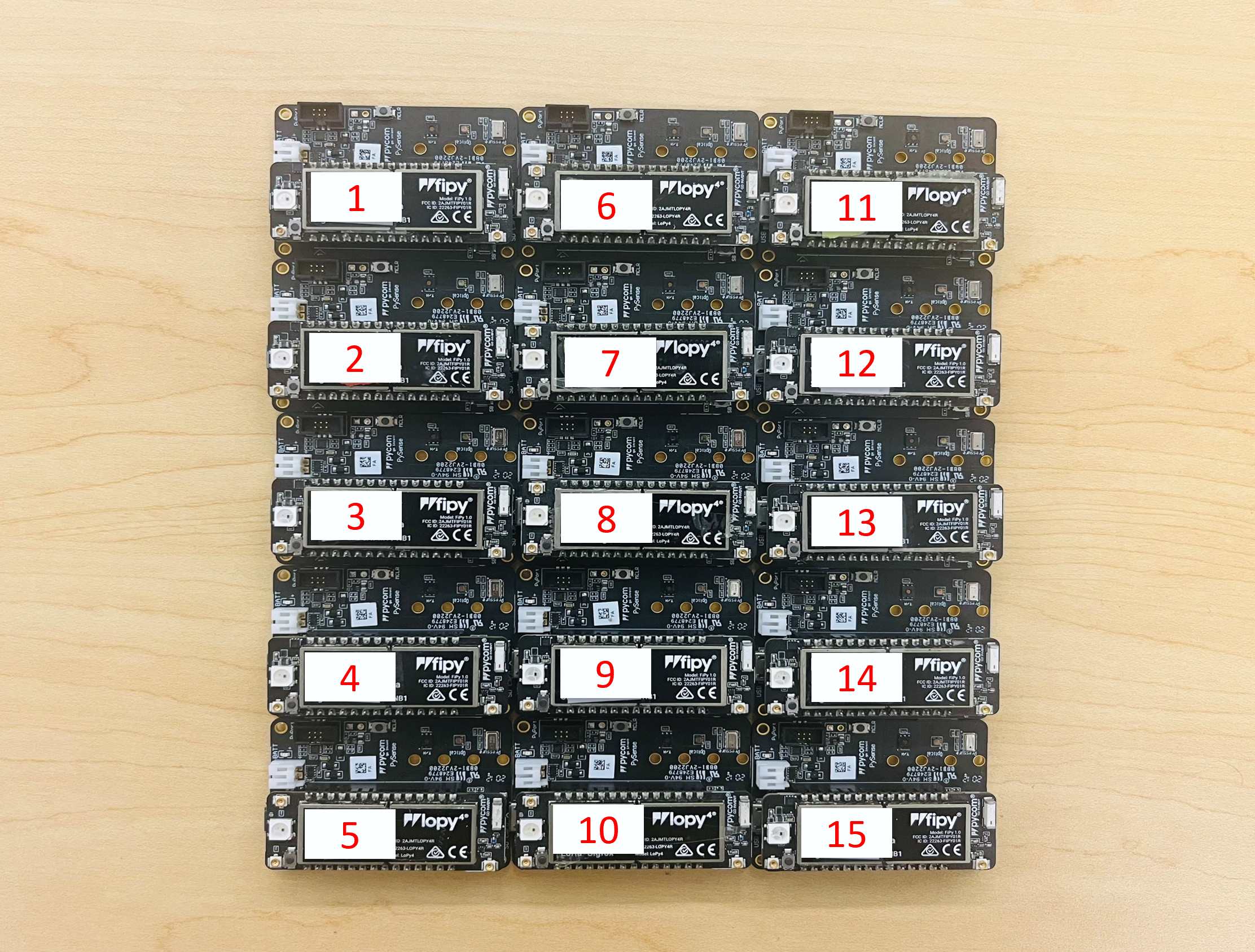}
     }
    \hspace{0.00001cm}
     \subfloat[Location Setup\label{loc_setup}]{%
       \includegraphics[width=0.23\textwidth, height = 0.2\textwidth]{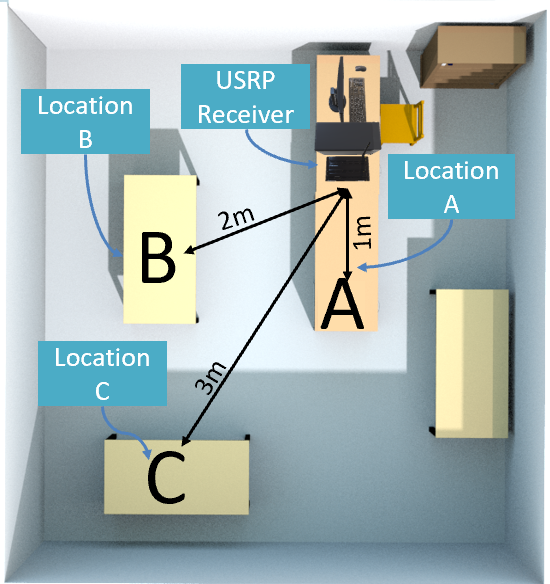}
     }
     \caption{IoT Testbed consisting of 15 Pycom transmitting devices and a USRP B210 receiving device}
     \label{testbed}
\end{figure}

To assess the effectiveness of \framework,
%
%
%
we considered two key performance metrics: same-domain accuracy and cross-domain accuracy. Same-domain accuracy measures the ability of the device identifier to identify devices accurately when the testing data/packets are drawn from the same training domain. On the other hand, cross-domain accuracy evaluates the models' ability to generalize across different domains, such as different locations, channels, or days. To ensure robust results, we utilized the 5-fold cross-validation technique, dividing each device's data into five non-overlapping partitions of equal size. {Furthermore, we compared the performance of our proposed \framework~with the same CNN framework fed with a typical IQ representation as an input (referred to as \iqcnn)}. 

\subsection{Device Identification Results and Analysis}
\subsubsection{Robustness to Fixed-Location Changes}\label{subsub:fix-loc}

\begin{figure*}
\centering
 \includegraphics[width=\linewidth]{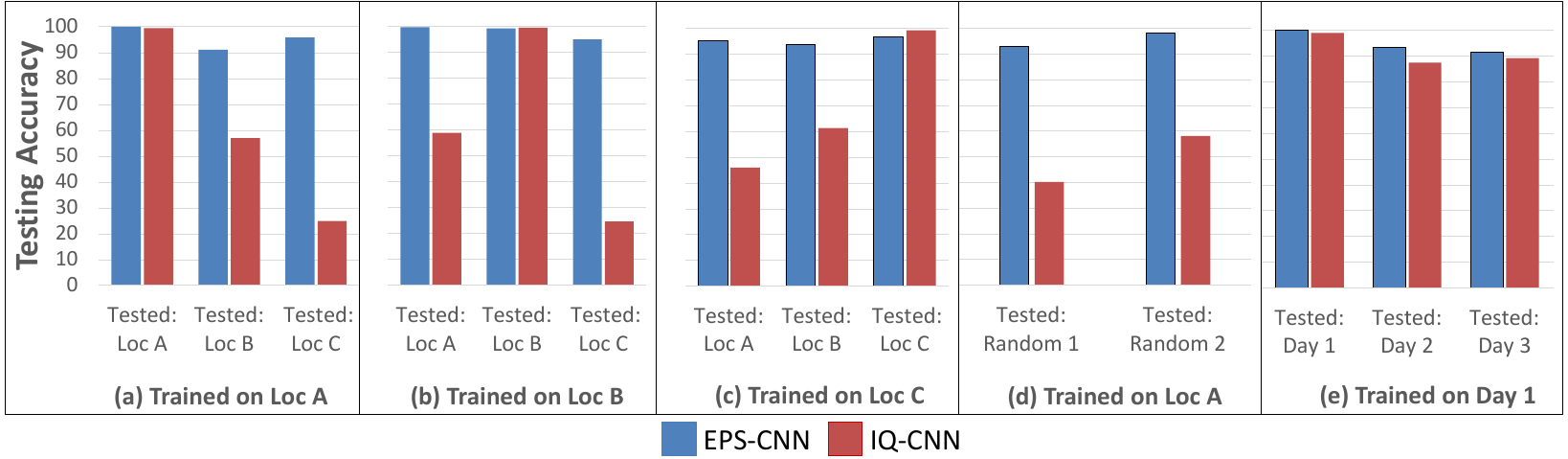}
 \caption{\framework~performance: the testing accuracies under the location and time scenarios} %
\label{eps-locations}
\end{figure*}


We begin by assessing the robustness of \framework~to changes in device locations, meaning that training and testing are done on data collected on different (but fixed) locations. For this, we leveraged our 15-device testbed, shown in Fig.~\ref{tx}, to collect WiFi 802.11b data for three different locations, Loc A, Loc B and Loc C, with devices being placed 1-meter, 2-meter and 3-meter away from the USRP receiver, respectively (see Fig.~\ref{loc_setup}).
%

Our findings, depicted in Fig.~\ref{eps-locations}(a)-(c), demonstrate remarkable  identification enhancements of \framework~over \iqcnn~in cross-location scenarios, where training and testing are done on different locations. For instance, when training is done in Loc A but testing is done in Loc C (see Fig.~\ref{eps-locations}(a)), the average testing identification accuracy achieved under \framework~exceeds 95\%, whereas that achieved under \iqcnn~is below 30\%. Similar significant enhancements are also seen when training is done in Loc B (Fig.~\ref{eps-locations}(b)) or Loc C (Fig.~\ref{eps-locations}(b)) but testing is done in different locations.  
To the best of our knowledge, this is the highest performance achieved by deep learning-based device fingerprinting methods when training and testing are done on different domains. 

In addition to improving robustness to location changes remarkably, \framework~does achieve exceptional testing accuracy when testing and training are done at the same location, whether Loc A, Loc B, or Loc C.
Observe that the average same-location testing accuracy achieved under \framework~at Loc A, Loc B, and Loc C are 100\%, 99.6\%, and 96.7\%, respectively. It is worth mentioning that in the case of same-location scenario, \iqcnn~too achieves high performances, as shown in Fig.~\ref{eps-locations}.


 

\subsubsection{Robustness to Random-Location Changes}\label{subsub:ran-loc}
We now turn our attention to evaluating the effectiveness of \framework~under random placement of devices. For this experiment, during training (also referred to as enrolment or registration), all devices transmit from a fixed location, 1m away from the receiver, but during testing, the devices transmit from random locations all within 3m from the USRP receiver. 
%
Fig.~\ref{eps-locations}d {shows that \framework~achieves high average cross-domain testing accuracies of 93\% and 98\%, respectively, when trained on Loc A and tested under two random-location deployments 1 and 2. In contrast, \iqcnn's performance deteriorates when tested under random-location placements, whose achieved average testing accuracies are only 40\% and 58\% under random-location deployments 1 and 2, respectively.} 

 \subsubsection{Robustness to Time Changes}
We now evaluate the resiliency of \framework~under the cross-day scenario, where training and testing are done on data collected on different days. For this, we collected WiFi datasets over three consecutive days, where all devices were placed 1 meter away from the receiver, and present the results of this experiment in Fig.~\ref{eps-locations}(e). 
The first observation we draw here is that {the same-day testing accuracies (both training data and testing data are collected on the same day) under \framework~and \iqcnn~were found to be 100\% and 99\%, respectively.}
More interestingly, {Fig.}~\ref{eps-locations}(e) shows that \framework~(and \iqcnn~to a lesser degree) maintains remarkable accuracy for the cross-day scenario, {where the average testing accuracies of \framework~are 93\% (compared to 88\% for \iqcnn) and 92\% (compared to 89\% for \iqcnn) when tested on Day 2 and Day 3 data, respectively. Similar significant enhancements were obtained when trained on Days 2 and 3 and tested on the other days.}

In recap, our findings shown in {Fig}.~\ref{eps-locations} {demonstrate the superiority of our \framework~framework compared to the conventional \iqcnn~framework in overcoming cross-location generalizability and adaptability.}

\subsection{Other Security Benefits}
{In addition to enhancing security, \framework~preserves privacy by minimizing the transmission of device credential information during network access operations. Moreover, it eliminates the need for error-prone manual upkeep of MAC Authentication Bypass/allow lists, as security policy rule recommendations are automatically derived from \framework's output.

The durability of our framework inherently depends on the aging rate of hardware components, especially the quality of the crystal oscillator, and the intra-distance between different fingerprints in the latent space. Consequently, it varies between deployments. In our current evaluation, we have concentrated on showcasing the resiliency of our framework against short-term aging, a standard practice in the assessment of newly proposed device fingerprinting techniques. This short-term analysis, conducted over a month, forms a robust foundation for understanding initial performance. However, we acknowledge the necessity to explore the model's behavior under prolonged aging circumstances, spanning months or even years, aligning with the typical lifespan of IoT devices.} 

%% file: 8a-conclusion.tex
{In conclusion, this article addresses security challenges posed by limitations in resource-constrained microcontroller-based devices, hindering their integration into the ZT security paradigm due to insufficient capabilities for robust identity-based access authentication. We present \framework, an innovative wireless device identification solution tailored for ZT architecture, emphasizing resource-constrained IoT devices. At its core is the unique \proposed~representation, overcoming domain challenges and establishing a robust foundation for device identity. Rigorous empirical validation demonstrates the practical viability of our approach, contributing to the advancement of secure IoT networks in an era marked by cyber threats and sophisticated attacks.}


%% file: main.bbl
\begin{thebibliography}{10}
\providecommand{\url}[1]{#1}
\csname url@samestyle\endcsname
\providecommand{\newblock}{\relax}
\providecommand{\bibinfo}[2]{#2}
\providecommand{\BIBentrySTDinterwordspacing}{\spaceskip=0pt\relax}
\providecommand{\BIBentryALTinterwordstretchfactor}{4}
\providecommand{\BIBentryALTinterwordspacing}{\spaceskip=\fontdimen2\font plus
\BIBentryALTinterwordstretchfactor\fontdimen3\font minus \fontdimen4\font\relax}
\providecommand{\BIBforeignlanguage}[2]{{%
\expandafter\ifx\csname l@#1\endcsname\relax
\typeout{** WARNING: IEEEtran.bst: No hyphenation pattern has been}%
\typeout{** loaded for the language `#1'. Using the pattern for}%
\typeout{** the default language instead.}%
\else
\language=\csname l@#1\endcsname
\fi
#2}}
\providecommand{\BIBdecl}{\relax}
\BIBdecl

\bibitem{ward2014beyondcorp}
R.~Ward and B.~Beyer, ``Beyondcorp: A new approach to enterprise security,'' 2014.

\bibitem{stafford2020zero}
V.~Stafford, ``Zero trust architecture,'' \emph{NIST special publication}, vol. 800, p. 207, 2020.

\bibitem{hunt2017seven}
G.~Hunt, G.~Letey, and E.~Nightingale, ``The seven properties of highly secure devices,'' \emph{tech. report MSR-TR-2017-16}, 2017.

\bibitem{sankhe2019oracle}
K.~Sankhe, M.~Belgiovine, F.~Zhou, S.~Riyaz, S.~Ioannidis, and K.~Chowdhury, ``Oracle: Optimized radio classification through convolutional neural networks,'' in \emph{IEEE INFOCOM 2019-IEEE Conference on Computer Communications}.\hskip 1em plus 0.5em minus 0.4em\relax IEEE, 2019, pp. 370--378.

\bibitem{jagannath2022comprehensive}
A.~Jagannath, J.~Jagannath, and P.~S. P.~V. Kumar, ``A comprehensive survey on radio frequency (rf) fingerprinting: Traditional approaches, deep learning, and open challenges,'' \emph{Computer Networks}, vol. 219, p. 109455, 2022.

\bibitem{elmaghbub2021}
A.~Elmaghbub and B.~Hamdaoui, ``{LoRa} device fingerprinting in the wild: Disclosing {RF} data-driven fingerprint sensitivity to deployment variability,'' \emph{IEEE Access}, 2021.

\bibitem{hussain2020machine}
F.~Hussain, R.~Hussain, S.~A. Hassan, and E.~Hossain, ``Machine learning in iot security: Current solutions and future challenges,'' \emph{IEEE Comm. Surveys \& Tutorials}, vol.~22, no.~3, pp. 1686--1721, 2020.

\bibitem{hamdaoui2022deep}
B.~Hamdaoui and A.~Elmaghbub, ``Deep-learning-based device fingerprinting for increased {LoRa-IoT} security: Sensitivity to network deployment changes,'' \emph{IEEE Network}, vol.~36, no.~3, pp. 204--210, 2022.

\bibitem{al2020exposing}
A.~Al-Shawabka, F.~Restuccia, S.~D’Oro, T.~Jian, B.~C. Rendon, N.~Soltani, J.~Dy, S.~Ioannidis, K.~Chowdhury, and T.~Melodia, ``Exposing the fingerprint: Dissecting the impact of the wireless channel on radio fingerprinting,'' in \emph{IEEE INFOCOM 2020-IEEE Conference on Computer Communications}.\hskip 1em plus 0.5em minus 0.4em\relax IEEE, 2020, pp. 646--655.

\bibitem{hanna_wisig_2022}
S.~Hanna, S.~Karunaratne, and D.~Cabric, ``Wisig: A large-scale wifi signal dataset for receiver and channel agnostic rf fingerprinting,'' \emph{IEEE Access}, vol.~10, p. 22808–22818, 2022.

\bibitem{needle}
A.~Elmaghbub and B.~Hamdaoui, ``A needle in a haystack: Distinguishable deep neural network features for domain-agnostic device fingerprinting,'' in \emph{2023 IEEE Conference on Communications and Network Security (CNS)}, 2023, pp. 1--9.

\bibitem{zhou2008frequency}
H.~Zhou, C.~Nicholls, T.~Kunz, and H.~Schwartz, ``Frequency accuracy \& stability dependencies of crystal oscillators,'' \emph{Carleton University, Systems and Computer Engineering, Technical Report SCE-08-12}, 2008.

\bibitem{vo2016fingerprinting}
T.~D. Vo-Huu, T.~D. Vo-Huu, and G.~Noubir, ``Fingerprinting wi-fi devices using software defined radios,'' in \emph{Proceedings of the 9th ACM Conference on Security \& Privacy in Wireless and Mobile Networks}, 2016, pp. 3--14.

\bibitem{elmaghbub2023eps}
A.~Elmaghbub and B.~Hamdaoui, ``{EPS}: distinguishable {IQ} data representation for domain-adaptation learning of device fingerprints,'' \emph{arXiv preprint arXiv:2308.04467}, 2023.

\bibitem{Park_McClellan}
T.~Parks and J.~McClellan, ``Chebyshev approximation for nonrecursive digital filters with linear phase,'' \emph{IEEE Transactions on Circuit Theory}, vol.~19, no.~2, pp. 189--194, 1972.

\end{thebibliography}
